# Fast three-dimensional phase retrieval in propagation-based x-ray tomography

**Darren Thompson**[ab*], **Yakov I. Nesterets**[ab], **Konstantin M. Pavlov**[cdb] and **Timur Gureyev**[eabdf]

[a] Commonwealth Scientific and Industrial Research Organisation, Clayton, Victoria, Australia

[b] University of New England, Armidale, New South Wales, Australia

[c] University of Canterbury, Christchurch, New Zealand

[d] Monash University, Clayton, Victoria, Australia

[e] ARC Centre of Excellence in Advanced Molecular Imaging, The University of Melbourne, Parkville, Victoria, Australia

[f] The University of Sydney, New South Wales, Australia

Correspondence email: darren.thompson@csiro.au

**Abstract** The following paper describes a method for three-dimensional (3D) reconstruction of multi-material objects based on propagation-based X-ray phase-contrast tomography (PB-CT) with phase retrieval using the homogenous form of the Transport-of-Intensity equation (TIE-Hom). Unlike conventional PB-CT algorithms that perform phase retrieval of individual projections, the described post-reconstruction phase-retrieval method is applied in 3D to a localised region of the CT-reconstructed volume. We demonstrate via numerical simulations the accuracy and noise characteristics of the method under a variety of experimental conditions, comparing it to both conventional absorption tomography and two-dimensional (2D) TIE-Hom phase retrieval applied to projection images. The results indicate that the 3D post-reconstruction method generally achieves a modest improvement in noise suppression over existing PB-CT methods. It is also shown that potentially large computational gains over projection-based phase retrieval for multi-material samples are possible. In particular, constraining phase retrieval to a localised 3D region of interest reduces the overall computational cost and eliminates the need for multiple CT reconstructions and global 2D phase retrieval operations for each material within the sample.

## 1. Introduction

X-ray imaging is a cornerstone of modern medical imaging with conventional two-dimensional (2D) radiography and three-dimensional (3D) computed tomography (CT) being common tools both in clinical and research domains (Bushberg *et al.*, 2012). Phase-contrast imaging (PCI) is a specialised modality where image contrast is achieved by exploiting both the refractive and absorption properties of the imaged object. This technique has been widely studied and refined for over fifty years since the pioneering work of Bonse and Hart (1965). PCI has become a powerful approach for improved imaging of soft-tissue samples, which often exhibit poor contrast-to-noise due to weak absorption contrast and photon fluence constraints in conventional absorption-based radiography (Bushberg *et al.*, 2012). With the ongoing refinement of synchrotron and micro-focus X-ray sources PCI is being adapted to medical use (Bravin *et al.*, 2013, Tromba *et al.*, 2016). Several PCI methods exist utilising different mechanisms to encode phase information into projection images. In common use are analyser-based imaging (ABI) (Goetz *et al.*, 1979, Gureyev & Wilkins, 1997, Davis *et al.*, 1995, Nesterets *et al.*, 2004) and grating-based imaging (GBI) (Cloetens *et al.*, 1997, Momose *et al.*, 2003, Weitkamp *et al.*, 2008, Nesterets & Wilkins, 2008) which utilise crystals and gratings, respectively, in the experimental setup. In this work we study the so-called propagation-based imaging (PBI), also known as in-line method (Snigirev *et al.*, 1995, Wilkins *et al.*, 1996). In contrast to the other PCI methods, exclusive of combined approaches (Pavlov *et al.*, 2004, Coan *et al.*, 2005, Pavlov *et al.*, 2005), PBI relies on the free-space propagation between the sample and detector for phase-contrast effects to manifest themselves as detectable intensity variations. Given this, PBI is simpler from an experimental perspective than other PCI methods. However, this simplicity is offset by more stringent requirements to spatial coherence properties of the incident X-ray beam (Nugent, 2010).

PBI has been shown to produce enhanced image contrast in weakly absorbing objects such as biological samples, generally in the form of diffraction fringes at the interfaces between different materials. However, material-specific quantitative information cannot be gleaned from PBI intensity images directly and requires the application of phase retrieval methods, prior to CT reconstruction, to recover the complex refractive index distribution $n(\boldsymbol{r}) = 1 - \delta(\boldsymbol{r}) + i\beta(\boldsymbol{r})$ within the sample. Several methods of phase retrieval have been developed in PBI, with different restrictions imposed on the object and imaging system. Transport of Intensity Equation (TIE) based methods for phase retrieval due to Teague (1983) and refined by others (Cloetens *et al.*, 1999, Bronnikov, 1999, Bronnikov, 2002) are commonly used. To reconstruct the 3D distribution of $n(\boldsymbol{r})$ most of these methods require multiple X-ray projections (at different propagation distances and/or X-ray energies) to be acquired at each angular position of the object, which can be difficult to achieve under experimental conditions with time and dose constraints.

A significant breakthrough was made by Paganin *et al.* (2002) who developed the so-called "homogeneous" TIE phase retrieval algorithm (TIE-Hom) that accurately reconstructs the complex refractive index of single-material or homogenous objects. The algorithm requires only a single projection for each view angle and is robust to noise. As such it has become the de-facto standard for phase retrieval in PB-CT. The TIE-Hom algorithm makes use of a spatially uniform "monomorphous" (Turner *et al.*, 2004) factor or delta-to-beta ratio, $\gamma = \delta / \beta$, which defines the relative weight of the phase shift and absorption in the material of interest and results in the simplification of the reconstruction of the complex refractive index to $n(\boldsymbol{r}) = 1 + (i - \gamma)\beta(\boldsymbol{r})$. In reality, most samples do not consist of a single material, so use of TIE-Hom phase retrieval requires a compromised choice of $\gamma$ for one particular material interface. Qualitatively this choice results in the blurring of edges at the interfaces of materials where $\gamma$ is overestimated and the retention of phase-contrast fringes for the underestimated case. Quantitatively, there are corresponding errors in the reconstructed distribution of $n(\mathbf{r})$. These errors have been shown to be reduced by the collection of additional projections or utilising suitable *a priori* information (Gureyev *et al.*, 2013).

The TIE-Hom method has also been extended to enable quantitatively accurate phase retrieval of images containing non-overlapping projections of two materials (Gureyev *et al.*, 2002) and subsequently of *m* materials (Beltran *et al.*, 2010). In both cases *a priori* information is available for values of $\gamma_m$ for each material interface. The method proposed by Beltran *et al.* (2010) demonstrates that a composite 3D reconstruction of $n(\mathbf{r})$ can be produced by *m* separate applications of TIE-Hom phase retrieval using different $\gamma_m$ to the projection set, each followed by CT reconstruction from which localised sub-volumes are spliced into the final reconstructed volume. The present work seeks to extend on this method for phase retrieval of multi-material samples, differing from the previously described approach whereby the material-specific TIE-Hom phase retrieval step is performed as a post CT-reconstruction filtering operation in 3D. Importantly, this 3D version of TIE-Hom is applied to localised sub-volumes, again using *a priori* values of $\gamma_m$ in each. The use of TIE-Hom phase retrieval discussed thus far has been applied to plane projections. This will now be referred to as pre-reconstruction 2D TIE-Hom (PreTIE-Hom2D). Multi-material phase retrieval has also been attempted by the application of a 3D correction filter operation applied to the reconstructed volume in addition to conventional 2D TIE-Hom (Ullherr & Zabler, 2015). Also, other non-TIE based methods of 3D phase retrieval have been developed (Vassholz *et al.*, 2016, Ruhlandt *et al.*, 2014, Maretzke *et al.*, 2016). For clarity our proposed method will be referred to as post-reconstruction 3D TIE-Hom (PostTIE-Hom3D) for which the derivation is given in section 2. A numerical simulation framework has been created to simulate, evaluate and compare these methods and is discussed in section 3.

## 2. Derivation – post-reconstruction 3D TIE-Hom phase retrieval

Consider an imaging system schematically shown in Fig.1. Let an object be illuminated by a monochromatic plane X-ray wave with wavelength $\lambda$ and intensity $I_{in}$, $I_{in}^{1/2} \exp(ikz)$ with $k = 2\pi/\lambda$. The image of the object is recorded on a position-sensitive detector located at the distance $R$ downstream from the object. In the following we assume that the dimensions of the object are small compared to the source-to-object distance $\rho$ and $\rho \gg R$. Interactions of the X-rays and object matter are described via the spatial distribution of the complex refractive index, $n(\mathbf{r}) = 1 - \delta(\mathbf{r}) + i\beta(\mathbf{r})$, $\mathbf{r} = (x, y, z)$.

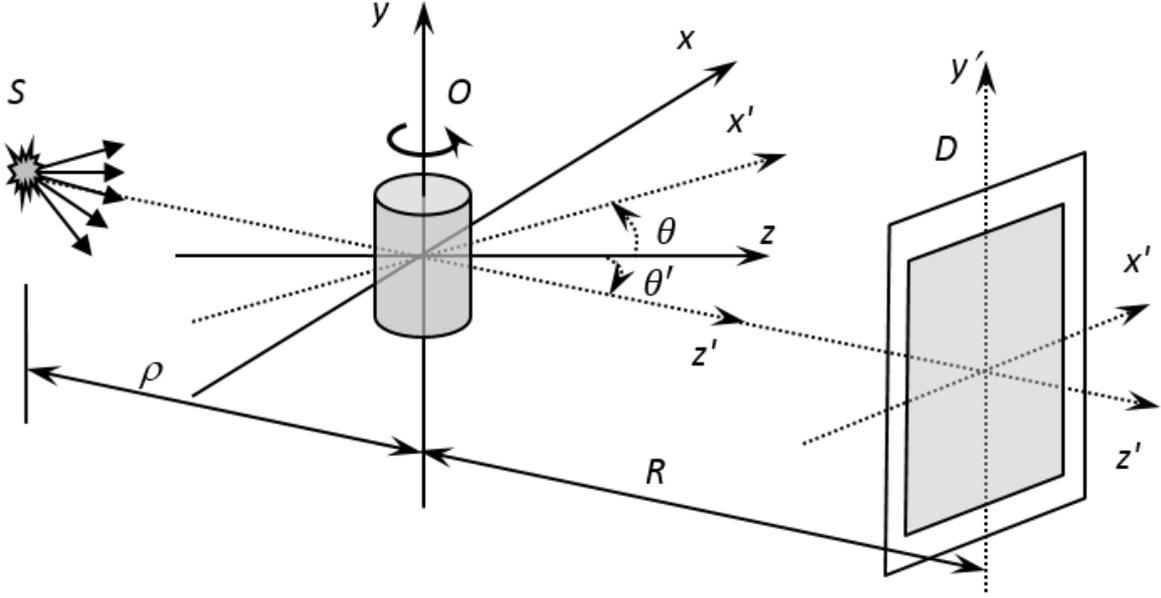

**Figure 1** PB-CT experimental setup, the direction of the incident X-ray wave forms an angle $\theta'$ with the z axis, $-\pi/2 \leq \theta' < \pi/2$, and $\theta = \theta' + \pi/2$, the object and detector planes are located at $z' = 0$ and $z' = R$ respectively.

If the projection approximation (Paganin, 2006) is applied to the PB-CT experimental setup depicted in Fig. 1, the following equations allow for the calculation of the transmitted phase and intensity respectively.

$$\varphi_\theta(x', y) = -k\mathbf{P}_\theta \delta(x', y), \tag{1}$$

$$I_\theta(x', y) = I_{in} \exp\{-2k\mathbf{P}_\theta \beta(x', y)\}, \tag{2}$$

where $\mathbf{P}_\theta f(x', y)$ represents the projection operator defined as:

$$\mathbf{P}_\theta f(x', y) = \int_{-\infty}^{\infty} \int_{-\infty}^{\infty} f(x, y, z) \delta_D(x' - x \sin\theta - z \cos\theta) dx dz. \tag{3}$$

Here $\delta_D(x)$ is the Dirac delta function.

The two-dimensional Fourier transform operator,

$$\mathbf{F}g(\xi', \eta) = \int_{-\infty}^{\infty} \int_{-\infty}^{\infty} \exp[i2\pi(x'\xi' + y\eta)] g(x', y) dx' dy, \tag{4}$$

combined with the so-called filtered back projection (FBP) operator, $\Re$ (Natterer, 2001),

$$\Re h(x, y, z) = \int_0^\pi \int_{-\infty}^{\infty} \int_{-\infty}^{\infty} \exp\{-i2\pi[\xi'(x \sin\theta + z \cos\theta) + \eta y]\} h(\xi', \eta, \theta) |\xi'| d\xi' d\eta d\theta, \tag{5}$$

allows the construction,

$$\Re \mathbf{F} \mathbf{P}_\theta f(x, y, z) = f(x, y, z). \tag{6}$$

This inversion equation forms the mathematical basis of CT, permitting the reconstruction of the 3D distribution $f(x, y, z)$ from a set of measured projections $P_\theta f(x', y)$ at angles $\theta$ within the interval $(0, \pi)$.

One can now utilise the FBP operator to obtain expressions to separately reconstruct the real and imaginary parts of the complex refractive index. By substituting rearrangements of Eq.(1) and (2) into Eq. (6) respectively we obtain the following pair of equations for $\beta$ and $\delta$:

$$\beta(x, y, z) = -(1/2k) \Re F \ln(I_\theta / I_{in})(x, y, z), \tag{7}$$

$$\delta(x, y, z) = -(1/k) \Re F \varphi_\theta(x, y, z). \tag{8}$$

Conventional X-ray radiography and CT are generally concerned with measuring the intensity distribution of transmitted radiation in the object plane and reconstructing the imaginary part of the complex refractive index, $\beta$, relating the absorption characteristics of the object to the measured intensity. We now seek to utilise the TIE to infer phase information from the visible diffraction fringes created upon propagation through a given distance. The finite-difference form of the TIE is given by:

$$I_\theta^R(x', y) = I_\theta(x', y) - (R/k) \nabla_\perp \cdot \left[ I_\theta(x', y) \nabla_\perp \varphi_\theta(x', y) \right], \tag{9}$$

where $\nabla_\perp = (\partial / \partial x', \partial / \partial y)$ and $I_\theta^R(x', y)$ is an in-line phase-contrast image in the detector plane. This method in general requires multiple (at least two) intensity measurements (at different $R$) in order to solve Eq. (9) for the unknown phase distribution (at a given $\lambda$), which may be difficult or problematic in the context of an experimental implementation. Assuming that the sample object is "monomorphous" such that a spatially-independent (but energy-dependent) proportionality constant, $\gamma = \delta / \beta$ holds for the complex refractive index (Paganin *et al.*, 2002, Mayo *et al.*, 2003). This assumption is valid, for example, for objects consisting of a single material and objects composed of light elements $(Z < 10)$ when irradiated with high-energy X-rays (60-500 keV) (Wu & Liu, 2005). Utilising this property, one can link phase and intensity by rearranging Eqs. (1) and (2) into

$$\varphi_\theta(x', y) = (\gamma / 2) \ln \left[ I_\theta(x', y) / I_{in} \right]. \tag{10}$$

Applying $\nabla_\perp$ to Eq. (10), to obtain:

$$\nabla_\perp \varphi_\theta(x', y) = (\gamma / 2) \left[ \nabla_\perp I_\theta(x', y) / I_\theta(x', y) \right]. \tag{11}$$

Then inserting Eq. (11) into the TIE, Eq. (9), to arrive at

$$I_\theta^R(x', y) = \left[ 1 - (\gamma R / 2k) \nabla_\perp^2 \right] I_\theta(x', y). \tag{12}$$

From the expression above, one notes that the intensity at $z' = R$ includes, in addition to the contact intensity, a phase-contrast term proportional to the 2D Laplacian of the contact intensity. A further simplification can be introduced if one considers the case of weak absorption whereby $2k\boldsymbol{P}_\theta\beta(x',y)\ll 1$ in Eq. (2):

$$I_\theta(x',y) \cong I_{in}\{1-(2k/\gamma)\boldsymbol{P}_\theta\delta(x',y)\}. \tag{13}$$

Inserting this approximation into Eq. (12) and rearranging:

$$1-I_\theta^R(x',y)/I_{in} = \left[(2k/\gamma)-R\nabla_\perp^2\right]\boldsymbol{P}_\theta\delta(x',y). \tag{14}$$

Moreover, since TIE implies weak phase contrast (Gureyev *et al.*, 2004) the approximation $1-I_\theta^R(x',y)/I_{in} \cong -\ln\left[I_\theta^R(x',y)/I_{in}\right]$ can be used. Let $K_\theta(x',y) = -\ln\left[I_\theta^R(x',y)/I_{in}\right]$ define the "in-line image contrast" function and using the Fourier space Laplacian identity, $\boldsymbol{F}\nabla_\perp^2 g(\xi',\eta) = -4\pi^2\left(\xi'^2+\eta^2\right)\boldsymbol{F}g(\xi',\eta)$, thus arriving at the Fourier transform of Eq. (14) :

$$\boldsymbol{F}\boldsymbol{P}_\theta\delta(\xi',\eta) = \boldsymbol{F}K_\theta(\xi',\eta)/\left[(2k/\gamma)+4R\pi^2\left(\xi'^2+\eta^2\right)\right]. \tag{15}$$

Inserting Eq. (15) into the FBP operator, Eq. (6) provides a "single-step" phase retrieval and CT reconstruction expression for monomorphous objects (Gureyev *et al.*, 2006, Arhatari *et al.*, 2007, Arhatari *et al.*, 2012),

$$\delta(x,y,z) = \Re\left(\boldsymbol{F}K_\theta(\xi',\eta)/\left[(2k/\gamma)+4R\pi^2\left(\xi'^2+\eta^2\right)\right]\right)(x,y,z). \tag{16}$$

A similar result has been derived by Bronnikov (Bronnikov, 1999, Bronnikov, 2002) for pure-phase objects with negligible absorption which can be seen to correspond to Eq. (16) with $\gamma \to \infty$. Similarly, Eq. (16) reduces to conventional pure absorption CT when $R=0$. Using an explicit form of the FBP operator, eq.(5), applying $(2k/\gamma)-R\nabla^2$, where $\nabla = (\partial/\partial x,\partial/\partial y,\partial/\partial z)$, to both sides of Eq.(16) and utilising the identity

$$\nabla^2 \exp\{-i2\pi[\xi'(x\sin\theta+z\cos\theta)+\eta y]\} = -4\pi^2(\xi'^2+\eta^2)\exp\{-i2\pi[\xi'(x\sin\theta+z\cos\theta)+\eta y]\},$$

the following expression for $\delta(x,y,z)$ is obtained,

$$\delta(x,y,z) = \left[(2k/\gamma)-R\nabla^2\right]^{-1}\Re\boldsymbol{F}K_\theta(x,y,z), \tag{17}$$

which concludes the derivation.

Equation (17) is the mathematical basis of our new approach for a 3D phase retrieval method for weakly absorbing monomorphous objects. As is seen, the phase retrieval step is decoupled from the FBP operator, implying that this step can be performed in 3D after the conventional CT reconstruction. A similar derivation for pure-phase objects is described in (Baillie *et al.*, 2012). One of the more interesting aspects to the form of Eq. (17) relates to the ability of performing phase retrieval localised to a region of interest (ROI). This property potentially gives an advantage over pre-reconstruction phase retrieval techniques for samples which may contain a range of different materials of interest whereby it is often difficult to optimize global parameters to achieve optimal contrast across the entire sample (Gureyev *et al.*, 2013). We will illustrate that with localisation we can apply a specific value of $\gamma$ chosen to "focus" on a desired material composition inside a 3D region, thus giving a quantitatively accurate phase retrieval within that region. A spatially localised form of phase retrieval also leads to some potential computational gains over existing methods, such as the ability to divide and parallelize phase retrieval over the sample reconstruction. The use of fast discrete Fourier or GPU based filters for implementing the 3D TIE-Hom filter on sub-regions would make this phase retrieval method computationally efficient, even for large datasets.

### 3. Numerical experiment setup

### 3.1. X-ray CT simulation framework

In order to evaluate the accuracy and characteristics of the PostTIE-Hom3D a computational simulation framework was constructed for conventional absorption CT and PB-CT workflows. This framework allows for the definition of an analytic 3D model consisting of multiple simple geometric primitives in space representing a material defined by its complex refractive index at a given X-ray energy. With such a model the workflow can be used to generate a volume image in addition to contact or PBI projections for a given number of rotation angles with specified resolution and photon statistics. These simulated projections can then be subsequently reconstructed, including phase retrieval as part of the processing pipeline and quantified with a range of metrics. The framework was primarily constructed using the ITK image processing toolkit (Ibáñez & Insight Software, 2005) and uses elements of the Astra-toolbox (van Aarle *et al.*, 2016) , X-TRACT (Gureyev *et al.*, 2011) and RTK (Rit *et al.*, 2014) packages for CT-reconstruction routines.

### 3.2. Multi-material numerical phantom

The numerical phantom model defined for these simulations consists of an air-filled $1024 \times 1024 \times 256\,\mu m^3$ cuboid containing a central $700\,\mu m$ diameter cylinder consisting of breast tissue with the composition described by Hammerstein *et al.* (1979) and serving as the background material. Embedded within the central cylinder are four smaller, $80\,\mu m$ diameter cylinders of different materials distributed azimuthally at a fixed distance from the origin. A schematic representation of the phantom and corresponding material properties for the objects at the X-ray energy of 20 keV are shown in Fig. 2 and Table 1 respectively. Values obtained for material specific $\beta$ and $\delta$ were calculated using the web-based tool (Gureyev *et al.*, 2011). The relative delta-to-beta ratio versus breast tissue, $\gamma_{rel}$ has been calculated for each material as $\gamma_{rel} = (\delta - \delta_{breast}) / (\beta - \beta_{breast})$, where $\delta_{breast}$ and $\beta_{breast}$ are the values for breast tissue.

**Table 1**  Material properties at *E*=20 keV

| Object | Material | $\beta \times 10^{10}$ | $\delta \times 10^7$ | $\gamma_{rel}$ (Breast tissue) |
|---|---|---|---|---|
| 1 | Air | 0.004641 | 0.00625 | 1681 |
| 2 | Breast Tissue | 3.48 | 5.85 | 1 |
| 3 | Weddellite (CaC$_2$O$_4$·2H$_2$O) | 35.17 | 10.36 | 141 |
| 4 | Paraffin (C$_{31}$H$_{64}$) | 1.93 | 5.35 | 322 |
| 5 | Adipose | 2.54 | 5.36 | 523 |
| 6 | Tumour | 3.57 | 5.93 | 950 |

The materials modelled within the inner cylinders include organic tumorous and adipose tissue. The calcite weddellite (CaC$_2$O$_4$·2H$_2$O, mass density 1.94 g/cm$^3$) represents one of the primary components of type I and type II breast calcifications (Ghammraoui & Popescu, 2017). Additionally, the hydrocarbon paraffin (C$_{31}$H$_{64}$, mass density 0.9 g/cm$^3$) was included as it can be used to enclose biopsy samples and has X-ray absorption and refractive properties similar to those of the other organic materials.

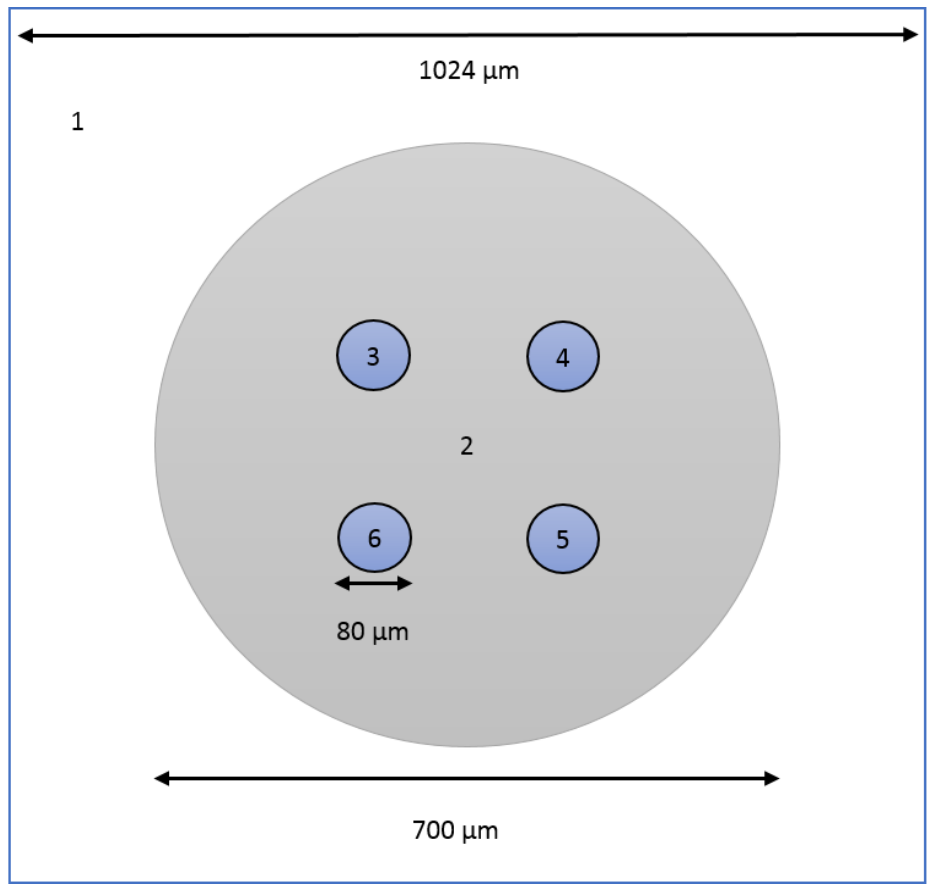

**Figure 2** Top view of the numerical phantom illustrating dimensions and materials as defined in Table 1.

### 3.3. Simulations

This research seeks to evaluate the numerical implementation of PostTIE-Hom3D derived in section 2 and compare its performance to conventional absorption CT and the Beltran *et al.* (2010) PreTIE-Hom2D method for the simulated multi-material phantom defined in section 3.2.

#### 3.3.1. Projection simulation

Distributions of transmitted intensity and phase shift from a simulated X-ray source (at 20 keV) were computed for $N_p$ sample rotation angles analytically from the phantom model definition in section 3.2 using Eqs. (1) and (2). Complex amplitude of the transmitted wave was then calculated over a 2D plane sampled at a given resolution. In the case of the simulations for this paper, a relatively fine sampling interval of 0.25 μm was chosen for the generation of initial projections. Due to the invariance of the phantom along the rotation *y*-axis, we can employ the simplification only requiring the generation of a single-row projection for the phantom at each angle step thus resulting in 1D plane-projection images of 8192x1 pixels.

For PBI projections, the Fresnel propagation operator (Paganin, 2006) implemented as a 2D Fourier filter was applied to each 2D projection, with the transfer function, $FG(\xi',\eta) = \exp[-i\pi\lambda R(\xi'^2 + \eta^2)]$. In the case of contact projections, the above propagation step was skipped leading to the final step of projection simulation where a discrete Gaussian smoothing filter was applied, (with the variance $\sigma^2 = 2$ μm$^2$) to simulate finite resolution of the detector. This step led to the smearing of sharp edges at the interfaces between objects and to reducing potential aliasing artefacts of subsequent phase retrieval and CT reconstruction. Finally, projections were sampled with a finite square aperture of 2 μm resulting in $N_p$ 1D row intensity projection images of 1024x1 pixels (32-bit real).

#### 3.3.2. Dose and noise

In this work we seek to model a fixed total radiation dose per scan utilising a specified exposure time per angle. To quantify the dose, we introduce the simulation parameter "total photons per pixel" (TPP) as the total number of incident photons per pixel per scan. In practice, we simulate noisy projections by applying a Poisson distribution with a known mean number of photons per individual pixel.

#### 3.3.3. Pre-reconstruction 2D TIE-Hom phase retrieval

For the application of the PreTIE-Hom2D method, a set of $N_p$ 2D projections was constructed by vertically stacking copies of the simulated, propagated and binned 1D row projections generated as per section 3.3.1. Poisson noise was then generated as described in the previous section, followed by

2D TIE-Hom phase retrieval, Eq. (18), applied to each 2D projection using the corresponding phantom material specific value of $\gamma_m$:

$$I_\theta(x', y) = \left[1 - \gamma_m R/(2k)\nabla_\perp^2\right]^{-1} I_\theta^R(x', y). \tag{18}$$

### 3.3.4. CT reconstruction

To recover the imaginary component $\beta$ of the complex refractive index from the thus far simulated projections the "gold-standard" FBP CT-reconstruction algorithm, Eq. (6) was applied. Prior to performing the actual reconstructions, standard background (flat-field) correction was performed before the $-\ln$ transform. In the case of the PreTIE-Hom2D multi-material method and as discussed in section 3.3.3, $N_p$ 2D phase-retrieved projections are reconstructed into a $1024^3$ voxel volume. To construct the final spliced multi-material volume, 2D TIE-Hom phase retrieval followed by FBP CT-reconstruction was performed for each phantom material separately with the composite volume constructed by inserting sub-volumes enclosing each cylinder. For PostTIE-Hom3D, $N_p$ 2D simulated projections were constructed following the approach described in section 3.3.3, without the phase retrieval step and then reconstructed with FBP producing a $1024^3$ voxel volume to which post-reconstruction 3D phase retrieval was applied.

### 3.3.5. Post-reconstruction 3D TIE-Hom phase retrieval

Successful application of PostTIE-Hom3D requires that the selected 3D region of interest (ROI) meets the following two criteria. Firstly, the ROI should fully contain the single material object under investigation. Ideally, this region should not be "polluted" with the inclusion of other objects or reconstruction artefacts which will lead to further undesirable artefacts in the phase-retrieved sub-volume. Secondly, the ROI should be chosen to consider the width of the 3D TIE-Hom point-spread function (PSF), $P_{TIE}$, which the Fourier transform of is defined as:

$$\boldsymbol{F}P_{TIE}(U) = \frac{1}{1 + AU^2}. \tag{19}$$

Here $U = (u^2 + v^2 + w^2)^{1/2}$, is the radial spatial frequency and $A = \pi\gamma\lambda R$ is a positive constant (we restrict our consideration to the case of positive $\gamma$). The corresponding real-space expression is

$$P_{TIE}(r) = \exp(-r/l_{TIE})/(4\pi l_{TIE}^2 r). \tag{20}$$

Here, $l_{TIE} = A^{\frac{1}{2}}/(2\pi)$ and $r = (x^2 + y^2 + z^2)^{1/2}$, with the standard deviation of this distribution equal to $\sqrt{6}l_{TIE}$. To overcome situations where the criteria are unable to be met with an appropriately large ROI due to neighbourhood constraints, it is possible to artificially enlarge the ROI by zero-padding to the required dimensions. Zero-padding to a power of two is also commonly applied in implementations of fast Fourier transform (FFT) based filters for optimal performance. For the simulations performed in this paper, the dimensions and locations of the cylindrical phantom objects are known explicitly, therefore a $128 \times 128 \times 128 \, \mu m^3$ cubic ROI was chosen for each of the four cylinders which adequately extends beyond the $80 \, \mu m$ cylinder diameter and easily satisfies the requirement for $P_{TIE}$. For example, given an X-ray energy of 20 keV, adipose tissue with $\gamma \cong 523$ (Table 1) and a propagation distance of $R = 60 \, mm$, results in the calculated characteristic length scale, $l_{TIE} \cong 12 \, \mu m$ corresponding to only several pixels at the simulated detector resolution.

### 3.3.6. Evaluation metrics

To quantify and compare the performance of the evaluated phase retrieval methods, two metrics have been selected, the Contrast-to-Noise ratio (CNR) (Gureyev *et al.*, 2014) and a Universal Image Quality Index (UIQI) introduced by Wang *et al.* (2002). CNR is defined as follows,

$$CNR = V^{1/2} \frac{\left| \langle \beta_o \rangle - \langle \beta_b \rangle \right|}{\left[ \left( \sigma_o^2 + \sigma_b^2 \right) / 2 \right]^{1/2}}, \tag{21}$$

where $V$ is the volume of the objects ROI (in voxels), $\langle \beta_o \rangle$ and $\sigma_o^2$ are the mean and variance of $\beta$ within the object ROI and $\langle \beta_b \rangle$ and $\sigma_b^2$ are the mean and variance for a "background" ROI devoid of any phantom objects. In these simulations, this background region corresponds to a similarly size region as the object ROI located outside all object ROI's in the reconstructed volume. With CNR we wish to compare the associated statistical gain between absorption only contact CT and the two phase retrieval methods which will highlight the respective noise suppression properties. As such, we calculate the CNR gain, $G_{CNR}$, which is simply the ratio of the CNR in the CT reconstructions using phase-retrieved and contact projection data, for same object and background ROI's:

$$G_{CNR} = \frac{CNR_{pr}}{CNR_{contact}}. \tag{22}$$

The second metric, UIQI,

$$UIQI = \frac{\sigma_{or}}{\sigma_o \sigma_r} \frac{2 \langle \beta_o \rangle \langle \beta_r \rangle}{\langle \beta_o \rangle^2 + \langle \beta_r \rangle^2} \frac{2 \sigma_o \sigma_r}{\sigma_o^2 + \sigma_r^2}, \tag{23}$$

requires a reference image for comparison, which is provided by the same ROI extracted from the numerical phantom model, with $\langle \beta_r \rangle$, $\sigma_r^2$ and $\sigma_{or}$ representing the mean, variance and covariance between the reference and sample ROI's. The UIQI produces a single numerical index that combines three factors: loss of correlation, luminance distortion and contrast distortion which its authors suggest permit the measurement of information loss as opposed to the quantification of error with other metrics (Wang & Bovik, 2002).

## 3.4. Results

Simulation parameters have been selected to be consistent with an experimental PB-CT imaging scenario representative of a small biological biopsy. For the simulations presented and discussed here, the following series of fixed parameters have been selected. An incident monochromatic X-ray beam with $E = 20$ keV and a pixel size, $h = 2$ μm (in both directions). A total of 900 projections were generated, $N_p = 900$ corresponding to an angular step of $0.2°$. This number conforms to the Nyquist sampling condition $N_p > (\pi/2)d/h$, for estimating the lower-bound for the number of required projections to avoid sub-sampling with $d$ being the reconstruction diameter (Hsieh & Spie, 2009). A propagation distance of $R = 60$ mm was chosen to be just below the limit for the TIE-Hom validity condition of $\lambda R / h^2 = 1$, such that good phase contrast should be obtained in the simulated propagated projections. A range of photon statistics were applied to the simulated projections as described in section 3.3.4. An exponentially increasing series of values in terms of the TPP and relative noise as a percentage were used, $1\times10^4$ (1.0%), $1\times10^5$ (~0.32%), $1\times10^6$ (0.1%), $1\times10^7$ (~0.032%), and $1\times10^8$ (0.01%). Notably, noise-free and contact projections were also simulated for comparative purposes.

The grid of 16 images presented in Fig. 3 comparatively illustrates PB-CT and the application of phase retrieval to FBP reconstructions of the simulated phantom for the imaging scenario described above in the case of 0.1% noise. Each image displays the central 2D slice through a $128\,\mu m^3$ cubic ROI of each material cylinder. Grid rows represents the materials: weddellite, paraffin, adipose and tumorous tissue; whilst columns display contact CT (R=0 mm), PB-CT (no phase retrieval), PreTIE-Hom2D and PostTIE-Hom3D for the propagation distance, R=60 mm. Visually inspecting the reconstructions for weddellite (Fig. 3a-d) some object contrast is achieved in the contact reconstruction (Fig. 3a) albeit with a reasonably high level of noise. Phase-contrast fringes are clearly visible in the reconstruction of the propagated case without phase retrieval (Fig. 3b) and similar high-quality results are shown for both TIE-Hom phase retrieval methods in Fig. 3c and d. The results for paraffin (Fig. 3e-f) and adipose tissue (Fig. 3i-l) are visually very similar, in both cases no apparent object contrast is perceptible in contact reconstructions. Again, noted is the presence of phase-contrast fringes in the non-phase retrieved images in Fig. 3f and 3j, although they are less defined and contain significantly more visible noise than the weddellite case. Both variants of phase retrieval produced visually similar results for the two materials and achieve relatively good contrast and reduced levels of noise such that the object is clearly distinguishable from the background. In the case of tumorous tissue (Fig. 3m-p) the results are far less revealing. It is worth noting that at $E = 20$ keV tumorous tissue has absorption and refractive properties more similar to the background breast tissue (Table 1) than the other materials in the phantom making the task of distinguishing the two regions after reconstruction significantly more difficult. This difficulty is clear in the non-phase retrieved images;

both contact and propagated images in Fig. 3m and 3n exhibit only noise. Notably, the success in the application of phase retrieval for tumorous tissue is not convincing when viewing a single reconstructed slice as in the case in Fig. 3o (PreTIE-Hom2D) and 3p (PostTIE-Hom3D) where little apparent contrast or structure is present for one to identify the reconstructed object.

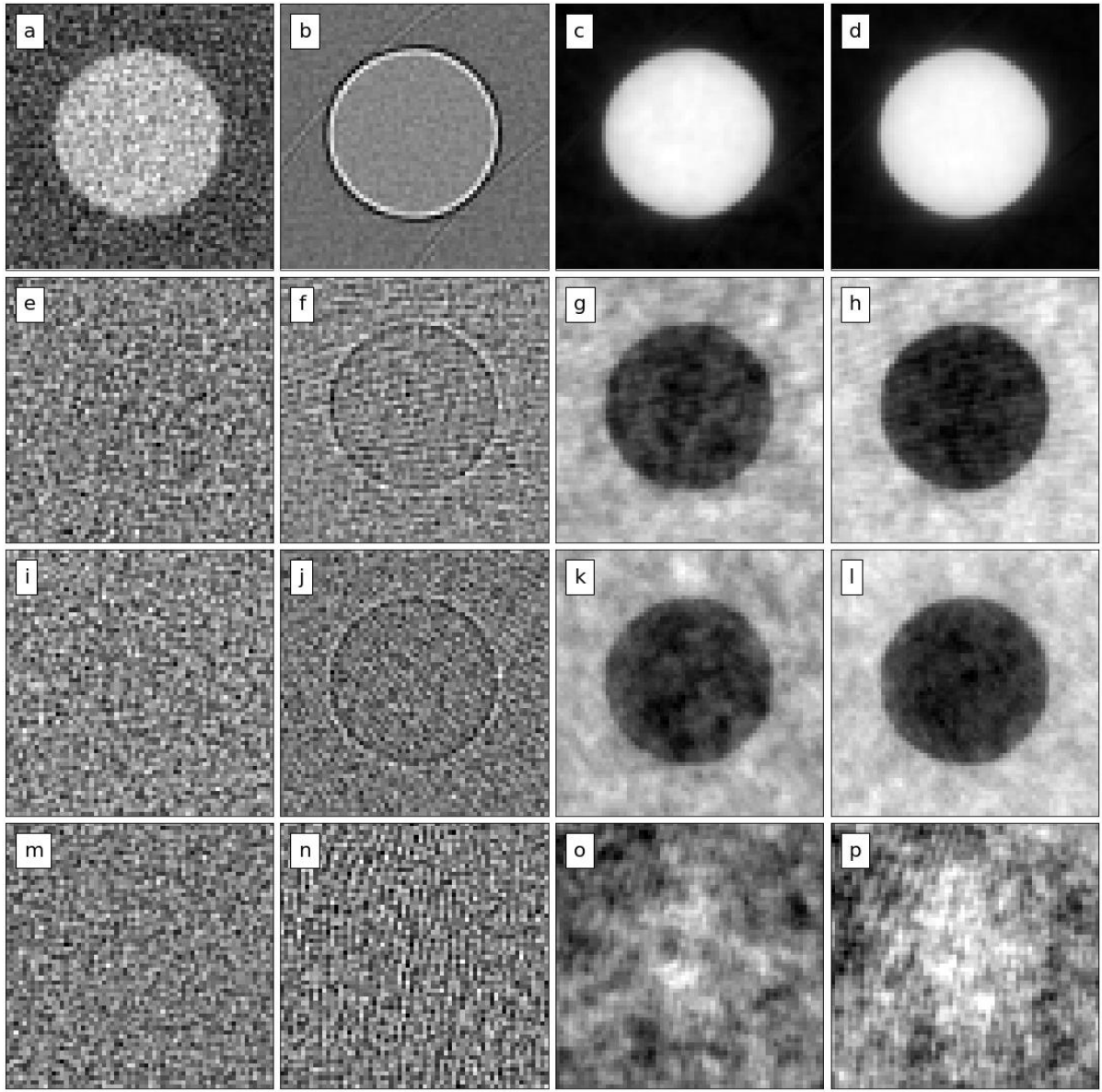

**Figure 3** Comparison of FBP reconstructions for the simulated phantom across the four materials with and without phase retrieval. Each image displays the central reconstructed slice in the volume ROI for the corresponding material with 0.1% noise. Rows display results for weddellite, paraffin, adipose and tumorous tissue respectively. Columns display reconstructions for contact-CT (R=0 mm), PB-CT (no phase retrieval), PreTIE-Hom2D and PostTIE-Hom3D (R=60 mm).

### 3.4.1. Reconstruction line-profiles

To gain insight into the relative quantitative merits of the two-phase retrieval methods, line profiles of the reconstructed central slice in each of the four materials regions of interest are presented in Fig. 5. Each plot presents the line profile of $\beta$ values over a central line extending 40 μm horizontally across the ROI. Separate plots for the reference computational model and reconstructed values after phase retrieval with PreTIE-Hom2D and PostTIE-Hom3D for $R = 60$ mm and 0.1% noise are plotted.

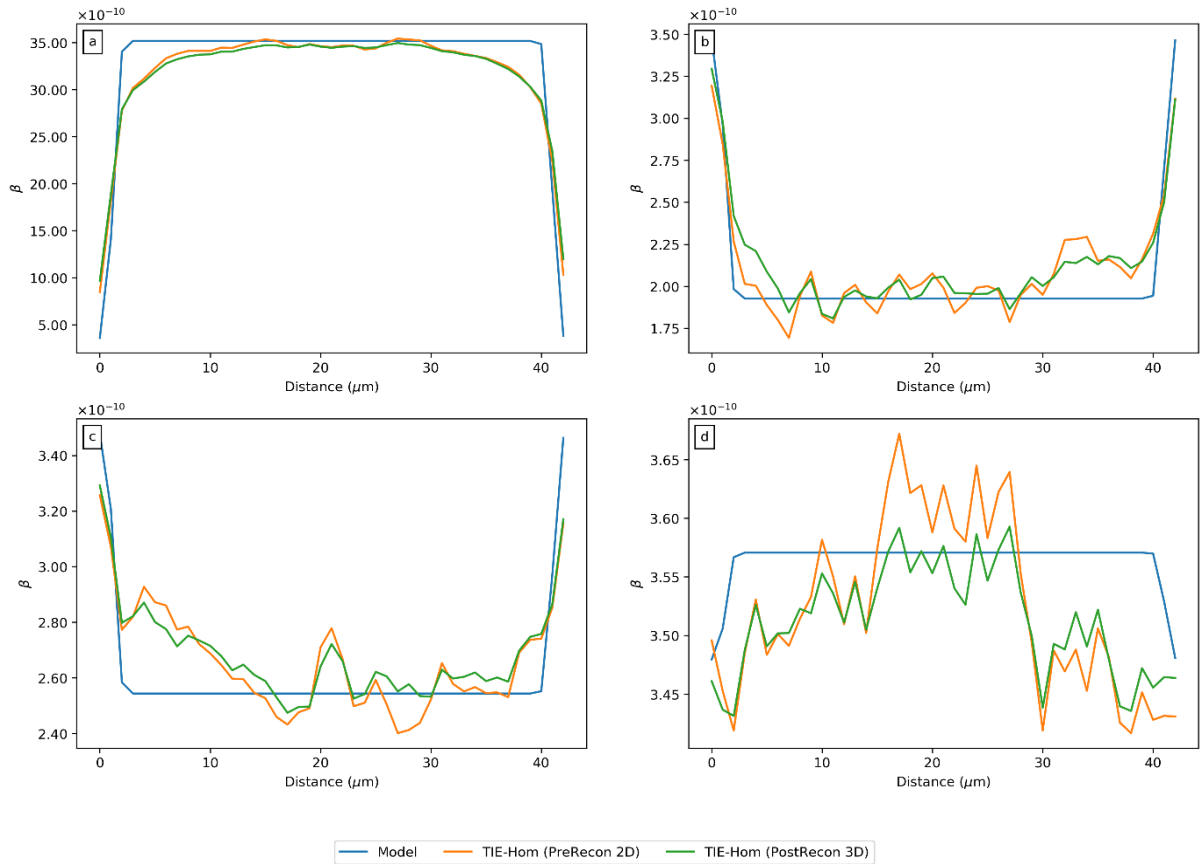

**Figure 4** Line-profile plots comparing the imaginary part of the complex refractive index, $\beta$ across the central reconstructed slice between the phantom model and PreTIE-Hom2D and PostTIE-Hom3D phase retrieval for $R = 60$ mm and 0.1% noise. (a) weddellite, (b) paraffin, (c) adipose and (d) tumorous tissue.

Inspection of Fig. 4a (weddellite) shows a similar reconstruction result with phase retrieval in both cases resulting in the elimination of diffraction fringes which are clearly visible in the contact reconstruction (Fig. 3c). Good contrast and suppression of noise have been achieved, giving rise to values of $\beta$ in agreement with the model. Notably, there is a similar degree of smoothing of the edges for both methods which may be attributed to the finite resolution of the imaging system and interpolation of projection data during CT reconstruction. The noise suppressing properties of PostTIE-Hom3D compared to PreTIE-Hom2D are illustrated for the less absorbing materials, paraffin (Fig. 4b) and Adipose tissue (Fig. 4c). Here, both profiles and images are similar, both materials exhibit comparable absorption and refractive properties at the simulated X-ray energy as evidenced by the values in Table 1. Again, for these simulation parameters, PostTIE-Hom3D displays qualitatively improved noise reduction for these materials. For the final material, tumorous tissue (Fig. 4d) the profiles indicate only limited success of both PreTIE-Hom2D and PostTIE-Hom3D in improving object contrast in this case. The object is barely perceptible after phase retrieval although it may be suggested that PostTIE-Hom3D is marginally less noisy.

### 3.4.2. Evaluation with respect to noise

The evaluation of the noise suppression, contrast enhancing properties and quantitative accuracy of both methods as displayed in the previous section needs to be viewed with respect to varying levels of Poisson noise. Figure 5 presents plots of $G_{CNR}$ and UIQI (section 3.3.6) for each material over a range of values of Poisson noise on a logarithmic scale. In the case of UIQI an additional plot for "contact" reconstructions, without any phase retrieval is displayed as a point of reference. The relative noise suppression characteristics between PreTIE-Hom2D and PostTIE-Hom3D are quantified by the calculated values of $G_{CNR}$ and UIQI respectively. Almost universally, both methods exhibit improved CNR at all noise levels for all four materials with PostTIE-Hom3D generally superior. However, some variation in the overall trends are observed across different materials. For example, in Fig. 5a (weddellite) and Fig. 5b (paraffin) we see that the reported values of $G_{CNR}$ are both maximal at the highest simulated noise level, TPP=$1\times10^4$ (1.0%), and decrease monotonically approaching 1 (no gain) at the lowest simulated noise level, TPP=$1\times10^8$ (0.01%). For adipose tissue (Fig. 5c), one notes quite different behaviour with both methods displaying an initial increase in gain, peaking at TPP=$1\times10^5$ (~0.32%) and then decreasing, approaching one at 0.01% noise. In this case it is also noted that PostTIE-Hom3D significantly outperforms PreTIE-Hom2D with respect to $G_{CNR}$ until their values align at around TPP=$1\times10^7$ (~0.032%). For tumorous tissue (Fig. 6d), one sees some similarity with that of adipose with general increase in $G_{CNR}$ with a peak shifted towards the lower-end of the simulated noise spectrum at TPP=$1\times10^7$ (~0.032%) and then falling away. Again, PostTIE-Hom3D exhibits a near uniform improvement in $G_{CNR}$ over PreTIE-Hom2D at noise levels approaching the peak after which both methods perform similarly, as previously noted.

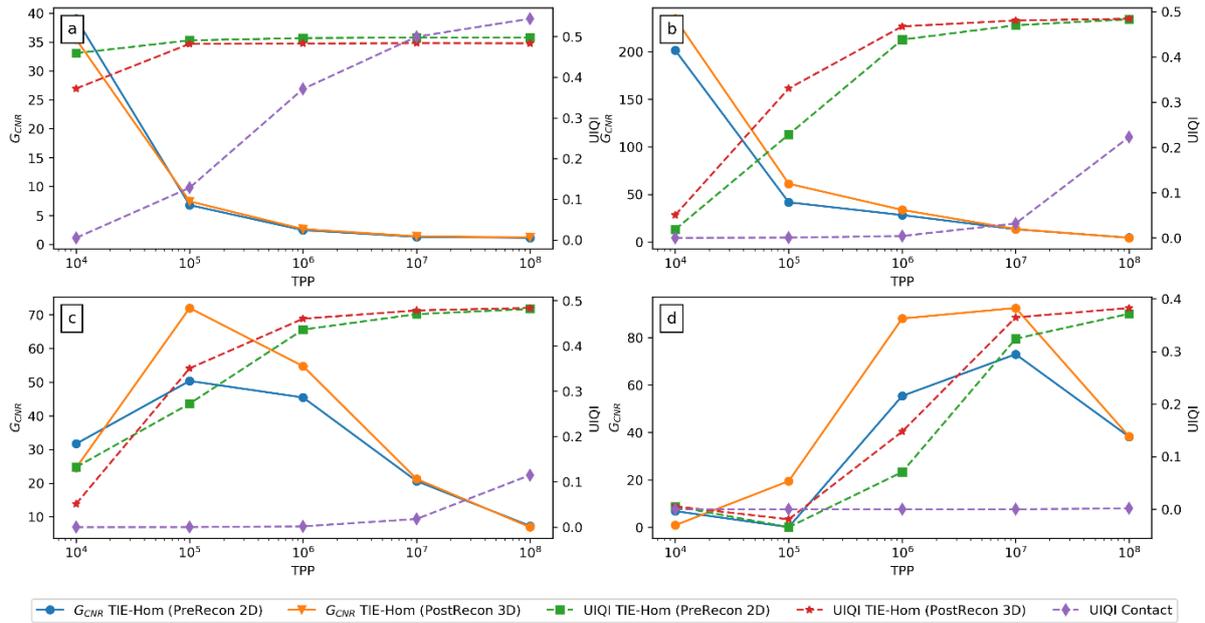

**Figure 5** Plots of $G_{CNR}$ and UIQI for $R = 60$ mm and varying levels of noise. (a) weddellite, (b) paraffin, (c) adipose tissue and (d) tumorous tissue.

Turning now to the plots for UIQI, as mentioned in the definition (section 3.3.6), this metric attempts to quantify "Image Quality" from a broader perspective than the purely statistical nature of $CNR$. Its use of a reference image in the form of the numerically accurate model ROI implies that it may provide values more representative of the quantitative deviation of the subject image from the reference. For two of the less absorbing materials, paraffin (Fig. 5b) and adipose (Fig. 5c) one observes very similar results with a monotonically increasing value of UIQI as TPP increases, with PostTIE-Hom3D giving a slightly higher quality result for higher noise levels until converging with PreTIE-Hom2D at around 0.01% noise. For these materials, one also notes that UIQI for "contact" reconstructions remains relatively uniform until around 0.032% noise from where the quality metric increases with corresponding decreased noise. With tumorous tissue (Fig. 5d) the UIQI for both phase retrieval methods dips below that of contact reconstructions before increasing in line with decreasing noise as with the other materials. This initial degradation in UIQI at higher noise levels for tumorous tissue can be attributed to the relatively little object information available to "retrieve" relative to noise such that phase retrieval introduces a greater deviation from the model. The more highly absorbing weddellite sample (Fig. 5a) illustrates different behaviour from the others in that UIQI for both phase retrieval methods remains relatively uniform and similar irrespective of the noise level. In contrast, UIQI for contact reconstructions of weddellite increases monotonically (as expected) with decreasing noise exceeding the corresponding values for both PostTIE-Hom3D and PreTIE-Hom2D for $TPP > 1 \times 10^7$.

The relative improvements in image quality due to phase retrieval in comparison to contact CT as presented quantitatively via $G_{CNR}$ and UIQI, as noise levels are varied, are shown in Fig. 6. Here, in a similar form as Fig. 3, central 2D FBP reconstructed slices for the ROI containing tumorous tissue are displayed for contact and PB-CT with phase retrieval for three different levels of noise (rows) corresponding to 0.1%, ~0.03% and 0.01%. Columns display contact-CT, PB-CT (no phase retrieval), PreTIE-Hom2D and PostTIE-Hom3D for the propagation distance, R=60 mm. Fig. 6a-d shows the same images as Fig 3m-p, corresponding to 0.1% noise. Reviewing the images for contact-CT in the first column for each noise level (Figs. 6a, 6e & 6j) one observes no perceivable contrast; all reconstructed slices appear to display uncorrelated noise. For the second column, PB-CT without phase retrieval one notes at ~0.03% (Fig. 6f) and 0.01% noise (Fig. 6j) the emergence of a diffraction fringe at the boundary interface between the tumorous and breast tissue in addition to reconstruction "streak" artefacts which are most likely due to sharp edges of the neighbouring weddellite object and are similar in appearance to those for adipose tissue and paraffin at the higher noise level of 0.1% as seen earlier in Figs. 3f & 3j. Turning now to the phase retrieval results in columns three and four, one sees that both PreTIE-Hom2D and PostTIE-Hom3D are able to produce results with enough contrast that the circular region of tumorous tissue is clearly distinguishable from the background breast tissue for ~0.03% noise (Figs. 6g-h) and slightly higher perceivable contrast at 0.01% noise (Figs. 6k-l). Again, noted is the appearance of streak artefacts and a marginally visually enhanced result in the case of PostTIE-Hom3D than PreTIE-Hom2D.

Overall, the results in this section illustrate the ability of TIE-Hom based phase retrieval in conjunction with PB-CT to reveal useable image contrast in the presence of noise where conventional contact-CT fails. Moreover, our research demonstrates that the level of contrast achieved is a function of the material, noise statistics and intrinsic properties of the imaging system and is consistent with theoretical derivations for the effect of phase retrieval in PB-CT (Nesterets & Gureyev, 2014).

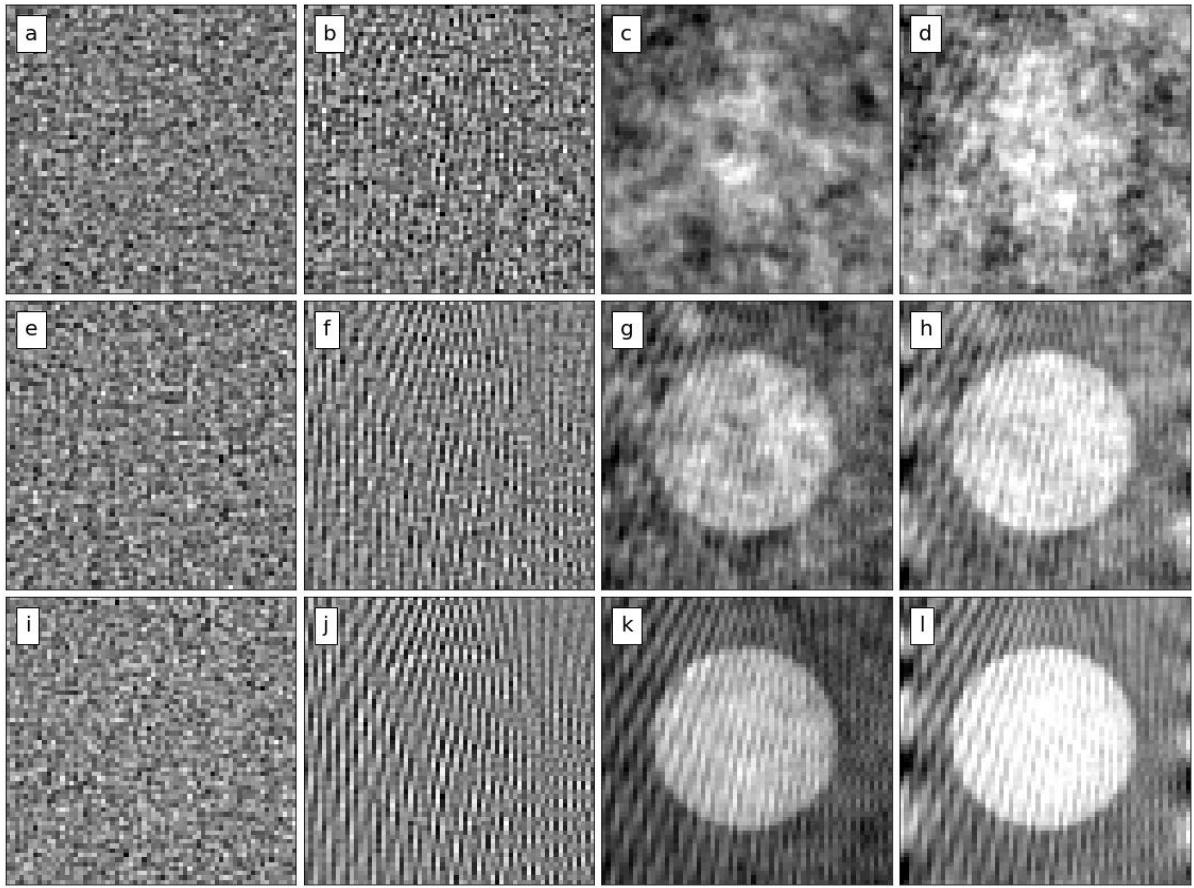

**Figure 6** Comparison of FBP reconstructions of the simulated phantom within the ROI containing tumorous tissue with and without phase retrieval and varying noise levels. Each image displays the central reconstructed slice in the volume ROI. Rows display results for noise levels of 0.1%, ~0.03% and 0.01%. Columns display reconstructions contact-CT (R=0 mm), PB-CT (no phase retrieval), PreTIE-Hom2D and PostTIE-Hom3D (R=60 mm).

### 3.4.3. Evaluation with respect to propagation distance

With some insight gained as to the characteristics of the phase retrieval schemes in the presence of varying levels of simulated noise and hence dose, the focus now becomes an evaluation of the response of varying propagation distance, which serves as the key experimental parameter in PB-CT. Several studies (Nesterets & Gureyev, 2014, Kitchen *et al.*, 2017) have demonstrated that increasing propagation distance in combination with TIE-based phase retrieval results in significantly improved signal-to-noise ratio in resulting images. Such improvements permit the use of a smaller radiation dose, producing images of similar quality to those produced without PB-CT methods at higher doses.

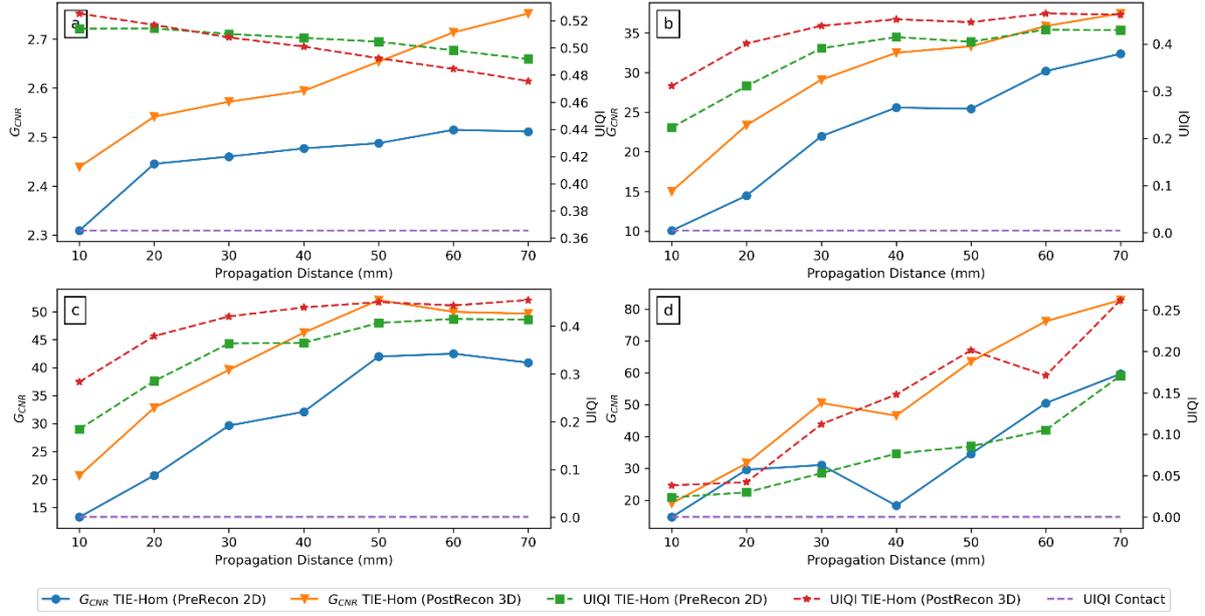

**Figure 7** Plots of $G_{CNR}$ and UIQI for $10 \text{ mm} \leq R \leq 70 \text{ mm}$ and TPP=$1 \times 10^6$. Plot (a) weddellite, (b) paraffin, (c) adipose tissue and (d) tumorous tissue.

For the simulations previously described at fixed propagation distance, $R = 60 \text{ mm}$ was used and corresponded to just below the distance for maximum phase contrast in the TIE regime defined by $\lambda R / h^2 = 1$. For the analysis in this section the noise level is fixed at 0.1%, TPP=$1 \times 10^6$ and the propagation distance varied with $10 \text{ mm} \leq R \leq 70 \text{ mm}$ at 10 mm increments. Figure 7 presents plots of $G_{CNR}$ and UIQI across this range of propagation distances. With respect to $G_{CNR}$ one sees relatively consistent trends across the four materials with gain generally increasing as propagation distance increases with PostTIE-Hom3D recording greater levels than PreTIE-Hom2D. The magnitude of the gain does vary with material with weddellite (Fig. 7a) achieving modest levels of the order $2.3 \leq G_{CNR} \leq 2.5$ for PreTIE-Hom2D and $2.4 \leq G_{CNR} \leq 2.8$ for PostTIE-Hom3D. The other less absorbing materials exhibit greater levels of $G_{CNR}$ with paraffin (Fig. 7b) exhibiting around $10 \leq G_{CNR} \leq 32$ (PreTIE-Hom2D), $15 \leq G_{CNR} \leq 36$ (PostTIE-Hom3D). Adipose tissue (Fig. 7c) reports slightly higher levels with $15 \leq G_{CNR} \leq 40$ for PreTIE-Hom2D and $20 \leq G_{CNR} \leq 50$ for PostTIE-Hom3D. In this particular case one also notes that there is a slight reduction in the magnitude of $G_{CNR}$ for propagation distances $R > 50 \text{ mm}$ which is the likely result of reconstruction artefacts beginning to dominate over the random noise. The results for tumorous tissue (Fig. 7d) differ again from previous cases with $G_{CNR}$ for both phase retrieval methods showing an initial rise then dip in the vicinity of $10 \text{ mm} \leq R \leq 40 \text{ mm}$ followed by a sustained increase for $R > 40 \text{ mm}$ where the magnitude for PreTIE-Hom2D is $15 \leq G_{CNR} \leq 60$ and $45 \leq G_{CNR} \leq 80$ for PostTIE-Hom3D. The graphs of $G_{CNR}$ over all four materials reveal that PostTIE-Hom3D consistently produces greater contrast over PreTIE-Hom2D by a relatively fixed, non-converging magnitude as the propagation distance increases.

Examination of the graphs for UIQI shows that again both paraffin (Fig. 7b) and adipose tissue (Fig. 7c) trend similarly with increasing propagation distance with nearly identical values in the range $0.2 \leq UIQI \leq 0.5$. As evident with $G_{CNR}$, PostTIE-Hom3D tends to outperform PreTIE-Hom2D by a fixed magnitude with UIQI levelling off from around $R \geq 60$ mm after which the theoretical validity conditions of TIE-Hom are exceeded, leading to a loss of fidelity in the phase retrieval process. Tumorous tissue (Fig. 7d) shows some inconsistency in the evaluated UIQI over both methods as was the case with $G_{CNR}$. This is likely due to the lack of available phase contrast in the projections to adequately reconstruct the object as seen visibly in the line profiles in Figure 5 and is quantified by the relatively low values of UIQI compared to the other materials. Despite this, PostTIE-Hom3D produces a UIQI nearly double that of PreTIE-Hom2D.

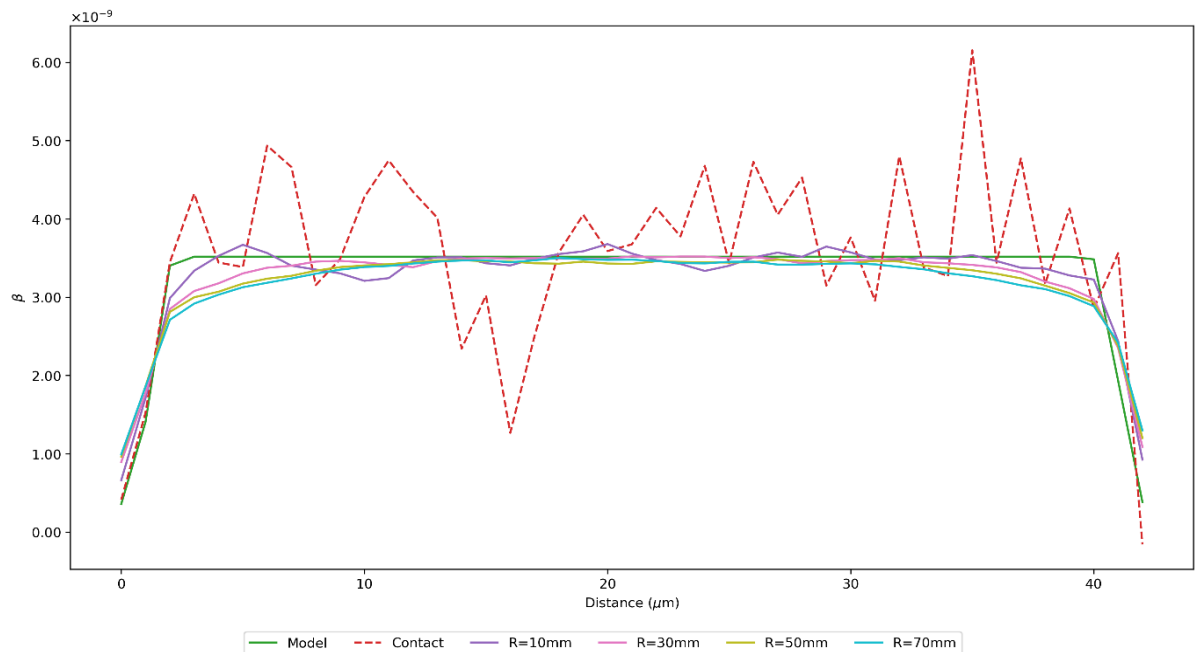

**Figure 8** Line profiles of $\beta$ values for the phantom model, contact and PostTIE-Hom3D phase retrieval for weddellite over propagations distances $10 \text{ mm} \leq R \leq 70 \text{ mm}$

The plot in Fig. 7a for weddellite shows distinctly different behaviour from the other materials. In this case one observes a monotonic decrease in UIQI for both methods as the propagation distance increases, opposite in behaviour to the others. Additionally, one notes that PreTIE-Hom2D outperforms PostTIE-Hom3D from around $R > 20$ mm where the latter decreases more rapidly as the propagation distance increases. Fig. 8 illustrates line profiles plotting PostTIE-Hom3D $\beta$ values for weddellite in the interval $10 \text{ mm} \leq R \leq 70 \text{ mm}$ with TPP=$1 \times 10^6$ in addition to the model and contact profiles. Evident from these plots is the ever-increasing degree of smoothing produced around the edges of the object as the propagation distance increases, resulting in an overall increased deviation from the reference model profile and degrading UIQI relative to the model as shown in Fig. 7a. Such over-smoothing can be attributed to the validity conditions of the TIE being exceeded for weddellite at increasing propagation distances.

Overall, the results display that the PB-CT in conjunction with the phase retrieval methods investigated result in significant improvements to image quality over contact-CT imaging. The results are consistent with those reported by Gureyev *et al*. (2014) illustrating improvements in contrast-to-noise of several orders of magnitude with increasing propagation distances. Moreover, such improvements correspond with an even greater relative reduction in the radiation dose required to obtain the equivalent quality to conventional contact imaging.

### 3.5. Computational analysis

Up to this point this paper has considered the merits of the two phase retrieval methods discussed purely in terms of their imaging characteristics. This section will now examine their performance from a computational perspective which is increasingly relevant in the context of real-life applications of PB-CT. Often, a realistic experimental scenario involves the use of synchrotrons or lab-based micro-CT systems leading to large datasets requiring dedicated high-performance computing (HPC) infrastructure to process within feasible timeframes. Such conditions lead to the need to consider the computation cost of the methods employed. Computationally, the application of the Beltran *et al.* (2010) multi-material method for a sample consisting of $N_m$ distinct materials over $N_p$ projections require $N_m \times N_p$ 2D phase retrieval operations in addition to $N_m$ FBP CT reconstructions. The computational complexity of the FBP algorithm is well studied (Natterer, 2001) and is dominated by the backprojection step. If one assumes optimal sampling conditions, the total work for reconstructing an $N_w^2 \times N_h$ volume is proportional to $N_h N_w^3$. TIE-Hom phase retrieval in 2D projection space is generally implemented as a 2D Fourier filter with total computational cost for $N_p$ projections proportional to $N_p N_w N_h \ln(N_w N_h)$. From these two components it is clearly seen that the dominant computational element of the Beltran *et al.* multi-material method is the $N_m$ FBP CT reconstructions, thus computationally bound by the number of materials to be examined and the sample dimensions, $N_m N_h N_w^3$. In contrast, the PostTIE-Hom3D method only requires a single FBP CT reconstruction with the application of phase retrieval implemented as a set of $N_o$ 3D Fourier filtering operations, where $N_o$ is the number of localised object regions of interest examined. The complexity of each 3D Fourier filter is proportional to $N_{ROI}^3 \ln(N_{ROI}^3)$ where $N_{ROI}^3$ is assumed to be a cube of uniform dimensions representing the region of interest. Therefore, the PostTIE-Hom3D method is computationally bounded by the number of objects examined and their size, $N_o$ and $N_{ROI}$, respectively. It is evident from these observations that PostTIE-Hom3D requires significantly less computation when both the number of objects and their size relative to overall sample under investigation is constrained. Conversely, the multi-material PreTIE-Hom2D method becomes computationally expensive as the number of materials examined increases. However, with modern HPC computing infrastructure a significant level of parallelization can be applied to the actual implementations of both methods. Utilising multi-core and GPU architectures it is feasible that all FBP reconstructions and phase retrieval steps could be executed in parallel. Such an implementation may result in both methods achieving comparative "temporal" performance, with the post-reconstruction method requiring significantly fewer computations overall.

## 4. Conclusion

Our research has derived a variant of TIE-Hom phase retrieval that can be applied directly to localised 3D regions of interest consisting of isolated single-material objects within a greater reconstructed volume of the distribution of the complex refractive index. Each region of interest requires *a priori* information relating to the absorption and refractive properties of the contained material. This method allows for efficient and accurate reconstruction of multi-material samples within the TIE regime with several marked benefits over alternative approaches. A simple numerical framework for PB-CT X-ray imaging has been constructed allowing for the method to be simulated and compared to the approach developed by Beltran *et al.* (2010) on a synthetic phantom under a range of experimental parameters. It is shown numerically using the contrast-to-noise and universal image quality index metrics that this new method achieves improved noise suppression, contrast enhancement and overall quantitatively correct results compared to contact absorption-only CT and TIE-Hom phase retrieval method applied to projections prior to CT reconstruction. It is also shown that the proposed 3D TIE-Hom method offers significant computational efficiencies for multi-material samples where performing post-reconstruction phase retrieval eliminates the need for multiple CT reconstructions, representing the most computationally expensive element. A potential practical application of this approach has been suggested by Paganin (2015) whereby PostTIE-Hom3D phase retrieval is applied to a conventional CT-reconstruction of phase-contrast projections, interactively adjusting the TIE-Hom gamma parameter and observing the result in order to obtain subjective "focus" for a specific material within the ROI. It is envisaged that such a method, if implemented in a software application using modern GPU hardware, could allow real-time combined reconstruction and visualization of multi-material objects.